\documentclass[spanish,english]{article}
\usepackage[T1]{fontenc}
\usepackage[latin9]{inputenc}
\usepackage{units}
\usepackage{amsmath}
\usepackage{amssymb}

\makeatletter

\newcommand*\LyXZeroWidthSpace{\hspace{0pt}}
\newcommand{\noun}[1]{\textsc{#1}}

\newcommand{\BA}{\begin{subequations}}
\newcommand{\CA}{\end{subequations}}
\usepackage{babel}

\usepackage{babel}

\usepackage{babel}

\usepackage{babel}
\addto\shorthandsspanish{\spanishdeactivate{~<>}}

\makeatother

\usepackage{babel}
\addto\shorthandsspanish{\spanishdeactivate{~<>}}

\begin{document}
\title{Solution of quantum eigenvalue problems by means of algebraic consistency
conditions}
\author{L. de la Peña, A. M. Cetto and A. Valdés-Hernández}

\maketitle
Instituto de Física, Universidad Nacional Autónoma de México,\\
 04510 Mexico
\begin{abstract}
We present a simple algebraic procedure that can be applied to solve
a range of quantum eigenvalue problems without the need to know the
solution of the Schrödinger equation. The procedure, presented with
a pedagogical purpose, is based on algebraic consistency conditions
that must be satisfied by the eigenvalues of a couple of operators
proper of the problem. These operators can be either bilinear forms
of the raising and lowering operators appropriate to the problem,
or else auxiliary operators constructed by resorting to the factorization
of the Hamiltonian. Different examples of important quantum-mechanical
textbook problems are worked out to exhibit the clarity and simplicity
of the algebraic procedure for determining the spectrum of eigenvalues
without knowing the eigenfunctions. For this reason the material presented
may be particularly useful for undergraduate students or young physicists. 
\end{abstract}

\section{Introduction}

One may find in the literature several approaches to the quantum eigenvalue
problem based on operator methods, the most well known ones being
the method of factorization of the Hamiltonian and the supersymmetric
approach\noun{ (susy}, see e.g. Refs. \cite{InfeldHull51,Wit81,Miel84,CooKha95}).
Reviews of \noun{susy} quantum mechanics from the point of view of
a general factorization method can be found in \cite{MiRO04,FeFe05};
further, Refs. \cite{CiPal77,CaPl09,JaTa13} contain examples of skillful
combinations of both approaches that are of relevance to the present
work.

On the other hand, since the early introduction of raising and lowering
operators for the harmonic oscillator and the angular momentum, significant
effort has been invested to study the relationship between ladder
operators and the factorization method. All in all, this relationship
has proved to be far from trivial. Specifically, it has become clear
that knowing the ladder operators does not mean that we know how to
factorize the Hamiltonian \cite{PerezL96}, and conversely, knowing
how to factorize the Hamiltonian does not in general lead to the construction
of differential ladder operators. The work by Infeld and Hull \cite{InfeldHull51},
followed by that of Mielnik and collaborators (see \cite{Miel84}
and references therein), has led first to a family of solvable potentials
isospectral to the harmonic oscillator, and thereafter to more general
solvable potentials with an arbitrary factorization energy. However,
as mentioned in \cite{Bermudez}, the factorization method is by now
well developed, and the set of problems that can be solved through
this technique can hardly be expanded. The procedure has also become
quite elaborate and difficult to follow for somebody not involved
in this area. A valuable general reference is the book by de Lange
and Raab \cite{deLange92}, which contains an accessible introduction
to operator methods for the solution of quantum mechanical problems.

Here we present an alternative method aimed at solving the quantum
eigenvalue problem without solving the Schrödinger equation, which
is characterized by its simplicity, generality and directness. The
proposed procedure, based on a few algebraic consistency conditions,
was first developed a number of years ago in Ref. \cite{PeMon80}
and applied to typical elementary textbook problems in which the ladder
operators do not depend on the state on which they operate, thereby
solving the eigenvalue problem without the need to know the solution
of the Schrödinger equation. Well-known instances are precisely the
harmonic oscillator and the angular momentum, as well as several other
textbook examples. A few years thereafter the algebraic technique
was successfully applied to solve the radial component of most popular
central-force quantum problems \cite{FeCa84}. Our purpose here is
to show how the proposed procedure can be generalized and applied
to a wider range of quantum-mechanical problems. Due to its transparency
and directness, the method of the consistency conditions could result
of particular importance or interest to undergraduate or early graduate
physics students.

For clarity in the presentation, the paper starts with an introduction
of the basic properties of ladder operators, which allows us to readily
derive some general algebraic consistency conditions that must be
satisfied by a couple of bilinear combinations of these operators.
The distinction is made between ladder operators with individual components
that either depend or do not depend on the state to which they are
applied. A couple of textbook examples illustrate the method for state-independent
ladder operators. It is further shown that the consistency conditions
can be applied also to solve quantum eigenvalue problems that lend
themselves to the factorization of the Hamiltonian, typically central-force
problems, including Dirac's H-atom. Further application to a simple
one-dimensional problem serves to stress the advantage of state-independent
operators as a tool to solve the eigenvalue problem by algebraic means.
Finally, an elementary example is presented that illustrates how the
algebraic method can also be used in connection with perturbative
calculations. 

\section{State-dependent and state-independent ladder operators}

Let us consider the eigenvector basis $\left\{ |k\rangle\right\} $
of a certain Hermitian operator $P$ of interest with discrete spectrum;
we use this basis to construct a pair of operators $\eta^{\dagger},\eta$
as follows, \begin{subequations} \label{eta}
\begin{equation}
\eta^{\dagger}=\sum_{i}C_{i}|i+1\rangle\langle i|=\sum_{i}C_{i}\eta_{i}^{\dagger},\label{13.122a}
\end{equation}
\begin{equation}
\eta=\sum_{i}C_{i}^{\ast}|i\rangle\langle i+1|=\sum_{i}C_{i}^{\ast}\eta_{i},\label{13.122b}
\end{equation}
with
\begin{equation}
\eta_{i}^{\dagger}=|i+1\rangle\langle i|,\;\eta_{i}=|i\rangle\langle i+1|\label{13.122c}
\end{equation}
\end{subequations} and with the constants $C_{i}$ to be determined
according to the needs. Note from \eqref{13.122c} that 
\begin{equation}
\eta_{i}^{\dagger}|n\rangle=|n+1\rangle\delta_{in},\;\eta_{i}|n\rangle=|n-1\rangle\delta_{i,n-1}.\label{13.122d}
\end{equation}
The individual operators $\eta_{i}^{\dagger},\eta_{i}$ may depend
explicitly or not on the state $i$ on which they operate, in which
case we speak of state-dependent or state-independent (individual)
operators, respectively. 

The action of $\eta^{\dagger}$ and $\eta$ given by \eqref{eta}
over a ket $|k\rangle$ of the basis is \begin{subequations} \label{etaket}
\begin{equation}
\eta^{\dagger}|k\rangle=\sum_{i}C_{i}|i+1\rangle\langle i|k\rangle=C_{k}|k+1\rangle,\label{13.123a}
\end{equation}
\begin{equation}
\eta|k\rangle=\sum_{i}C_{i-1}^{\ast}|i-1\rangle\langle i|k\rangle=C_{k-1}^{\ast}|k-1\rangle.\label{13.123b}
\end{equation}
\end{subequations} so that $\eta^{\dagger}$ and $\eta$ are, respectively,
raising and lowering operators in the Hilbert space of $\left\{ |k\rangle\right\} $,
for any selection of the coefficients $C_{k}$. It also follows that
\begin{subequations}
\begin{equation}
\eta\eta^{\dagger}|k\rangle=|C_{k}|^{2}|k\rangle,\label{13.124a}
\end{equation}
\begin{equation}
\eta^{\dagger}\eta|k\rangle=|C_{k-1}|^{2}|k\rangle.\label{13.124b}
\end{equation}
\end{subequations} We see that the basis $\{|k\rangle\}$ corresponds
to the eigenvectors of the operators $\eta\eta^{\dagger}$ and $\eta^{\dagger}\eta$,
and the quantities $|C_{k}|^{2}$, $|C_{k-1}|^{2}$ are the corresponding
eigenvalues. Therefore, the operators \label{A,S} 
\begin{equation}
S\equiv\eta\eta^{\dagger}+\eta^{\dagger}\eta=\{\eta,\eta^{\dagger}\},\quad A\equiv\eta\eta^{\dagger}-\eta^{\dagger}\eta=[\eta,\eta^{\dagger}],\label{SA}
\end{equation}
which are respectively the anticommutator and commutator of $\eta$
and $\eta^{\dagger}$, satisfy the eigenvalue equations 
\begin{equation}
S\left|k\right\rangle =s_{k}\left|k\right\rangle ,\;A\left|k\right\rangle =a_{k}\left|k\right\rangle ,\label{eq:skak}
\end{equation}
with \begin{subequations} \label{skak} 
\begin{equation}
s_{k}=\left|C_{k}\right|^{2}+\left|C_{k-1}\right|^{2},\label{sk}
\end{equation}
\begin{equation}
a_{k}=\left|C_{k}\right|^{2}-\left|C_{k-1}\right|^{2},\label{ak}
\end{equation}
\end{subequations} whence 
\begin{equation}
2\left|C_{k}\right|^{2}=s_{k}+a_{k}=s_{k+1}-a_{k+1}\label{8}
\end{equation}
and 
\begin{equation}
s_{k+1}-s_{k}=a_{k+1}+a_{k}.\label{consis}
\end{equation}
These basic consistency conditions between the eigenvalues of $S$
and $A$ corresponding to adjacent states will play an important role
in what follows. In particular, it is clear from Eqs. \eqref{skak}
that 
\begin{equation}
s_{k}\geq|a_{k}|.\label{10}
\end{equation}
We will consider that the spectrum of interest is bounded from below.
As is customary, we shall denote the ground state with $\left|k_{\min}\right\rangle =\left|0\right\rangle $
unless otherwise specified, so that, from Eq. \eqref{13.123b}, 
\begin{equation}
C_{-1}=0,\label{Ckmin}
\end{equation}
and from Eqs. (\ref{skak}), 
\begin{equation}
s_{0}=a_{0}.\label{min}
\end{equation}
 Successive application of Eq. (\ref{consis}), together with (\ref{min}),
leads to the recurrence relation 
\begin{equation}
s_{k}=a_{k}+2{\displaystyle \sum_{i=0}^{k-1}(-1)^{i-(k-1)}s_{i},}\label{rec}
\end{equation}
establishing the relationship between the eigenvalues of $A$ and
those of $S$.

In the case of a finite Hilbert space, with $\{\left|k\right\rangle \}=\{\left|0\right\rangle ,\dots,\left|k_{\max}\right\rangle \}$,
the upper bound imposes the condition $\eta_{k}^{\dagger}\left|k_{\max}\right\rangle =0$,
so that $C_{k_{\max}}=0$ and 
\begin{equation}
s_{k_{\max}}=-a_{k_{\max}}.\label{max}
\end{equation}

As we shall see below, for some applications it proves convenient
to introduce a couple of auxiliary operators $\alpha,\beta$, indirectly
defined through the relations \begin{subequations}\label{alfabeta}
\begin{equation}
\eta^{\dagger}=\tfrac{1}{\sqrt{2}}(\alpha-i\beta),\label{13.132a}
\end{equation}
\begin{equation}
\eta=\tfrac{1}{\sqrt{2}}(\alpha+i\beta).\label{13.132b}
\end{equation}
\end{subequations} From Eqs. \eqref{SA} and \eqref{alfabeta} it
follows that \begin{subequations}\label{betaalfa} 
\begin{equation}
A=i[\beta,\alpha],\label{13.133a}
\end{equation}
\begin{equation}
S={\alpha}^{2}+{\beta}^{2}.\label{13.133b}
\end{equation}
\end{subequations} In this form, when the ladder operators are known,
the consistency conditions allow us to determine the spectrum of $A$
in terms of the spectrum of $S$ or vice versa. This is particularly
useful when either one of these operators is identified with a dynamical
variable of interest, as is the case in the examples discussed in
Section 3.1.

It may happen that the operator corresponding to the relevant dynamical
variable---the one whose eigenvalues are looked for---is not of
the form of $A$ or $S$, but a linear combination of these (with
$p_{0}$, $p_{S}$ and $p_{A}$ real parameters) \cite{CaPl09,PeMon80},
\begin{equation}
P=p_{0}\mathbb{I}+p_{S}S+p_{A}A.\label{p}
\end{equation}
As exemplified in Section 3.2, the algebraic method may be applied
in some cases even without resorting to the explicit form of the ladder
operators, by establishing a functional relationship between the operators
$A$ and $S$ through an auxiliary operator $B$.

Before closing this section, we note that the definition of ladder
operators given in Eqs. \eqref{eta} can be readily extended to operators
connecting states that are not adjacent but separated a distance $T$,
i. e.,

\[
\eta^{(T)}\left|k\right\rangle =C_{k-T}^{(T)}\left|k-T\right\rangle .
\]
The corresponding operators $S^{(T)}$, $A^{(T)}$ satisfy the eigenvalue
equations 
\[
S^{(T)}\left|k\right\rangle =s_{k}^{(T)}\left|k\right\rangle ,\;A^{(T)}\left|k\right\rangle =a_{k}^{(T)}\left|k\right\rangle ,
\]
leading to 
\[
2\left|C_{k}^{(T)}\right|^{2}=s_{k}^{(T)}+a_{k}^{(T)}=s_{k+T}^{(T)}-a_{k+T}^{(T)}
\]
and 
\[
s_{k+T}^{(T)}-s_{k}^{(T)}=a_{k+T}^{(T)}+a_{k}^{(T)},
\]
as immediate generalizations of Eqs. \eqref{8} and \eqref{consis},
respectively, with $C_{-T}^{(T)}=0$. It is important to notice, however,
that although $T$ subsequent operations with single-step raising
operators take the system from state $\left|k\right\rangle $ to state
$\left|k+T\right\rangle $, the combined coefficient resulting from
such operations is not equal to $C_{k}^{(T)*}$ in general, as the
reader may confirm from the examples presented in the following sections.

\section{Applications}

\subsection{Textbook examples}

\subsubsection{\textit{1-D harmonic oscillator}}

As a first and most useful example of state-independent ladder operators
we recall the well-known case of the one-dimensional harmonic oscillator
(HO) of mass $m$ and frequency $\omega$, with raising and lowering
operators given respectively by
\begin{equation}
\eta^{\dagger}=\frac{1}{2\sqrt{m}}\left(m\omega x-ip\right),\;\eta=\frac{1}{2\sqrt{m}}\left(m\omega x+ip\right),\label{134a'}
\end{equation}
with $p=-i\hbar(d/dx)$. Notice that the ladder operators $\eta^{\dagger},\eta$
differ from the standard ones $a^{\dagger},a$ by a factor $\sqrt{\hbar\omega/2}$.
From Eqs. \eqref{alfabeta} we get 
\begin{equation}
\alpha=\sqrt{\frac{m}{2}}\omega x,\quad\beta=\frac{1}{\sqrt{2m}}p,\label{134a}
\end{equation}
which introduced in Eqs. \eqref{betaalfa} gives 
\begin{equation}
A=-\frac{i\omega}{2}\left[x,p\right]=\frac{1}{2}\hslash\omega,\qquad S=\frac{1}{2m}(m^{2}\omega^{2}x^{2}+p^{2})=H.\label{134b}
\end{equation}
The corresponding eigenvalues are therefore 
\begin{equation}
a_{k}=\frac{1}{2}\hslash\omega,\qquad s_{k}=E_{k}.\label{13.134c}
\end{equation}
Equation (\ref{10}) and the consistency condition \eqref{consis}
give 

\[
E_{k}\geq\frac{1}{2}\hslash\omega,\qquad E_{k+1}-E_{k}=\hslash\omega.
\]
Iterating the last expression gives 
\begin{equation}
E_{k}=k\hslash\omega+E_{0},\;\;\;k=0,1,...,\label{eq:60.3}
\end{equation}
where $E_{0}$ corresponds to the ground state, $s_{0}=a_{0}=\frac{1}{2}\hslash\omega$.
For a detailed history of the theory of the quantum harmonic oscillator,
see Ref. \cite{RuFr21}.

\subsubsection{\textit{Angular momentum}}

This is another important example of state-independent ladder operators\foreignlanguage{spanish}{
that is directly amenable to the above procedure. The angular-momentum
operators $\mathbf{J}=(J_{1},J_{2},J_{3})$ are defined as usual through
the eigenvalue equations }\begin{subequations} \label{jota} \foreignlanguage{spanish}{
\[
\mathbf{J}^{2}|\lambda,\mu\rangle=\lambda|\lambda,\mu\rangle,
\]
\[
J_{3}|\lambda,\mu\rangle=\mu|\lambda,\mu\rangle,
\]
}and the commutation relation (in units of $\hbar$),

\selectlanguage{spanish}%
\[
J_{3}=-i[J_{1},J_{2}].
\]
\foreignlanguage{english}{\end{subequations} }From these, it is a
simple exercise to construct the raising and lowering operators 
\begin{equation}
\eta^{\dagger}=\tfrac{1}{\sqrt{2}}(J_{1}+iJ_{2})=J_{\text{+}},\quad\eta=\tfrac{1}{\sqrt{2}}(J_{1}-iJ_{2})=J_{-}.\label{20.1}
\end{equation}
Since $J_{+},$ $J_{-}$ and $J_{3}$ commute with $\mathbf{J}^{2}$,
they leave the value of $\lambda$ unchanged and operate on $\mu$
only. This means that we look for the eigenvalues of $J_{3}$ for
a given magnitude of the total angular momentum (i. e., for fixed
$\lambda$). The construction of ladder operators for orbital angular
momentum that modify the value of $\lambda$ can be seen in \cite{LaRa86}.

Equation \eqref{20.1} suggests introducing 
\[
\alpha=J_{1},\quad\beta=-J_{2},
\]
so that from \foreignlanguage{english}{Eqs. \eqref{betaalfa}} 
\begin{equation}
A=i[J_{1},J_{2}]=-J_{3},\label{20.2-1}
\end{equation}
\begin{equation}
S=J_{1}^{2}+J_{2}^{2}=\mathbf{J}^{2}-J_{3}^{2},\label{20.3}
\end{equation}
whence 
\begin{equation}
a_{k}=-\mu_{k},\quad s_{k}=\lambda-\mu_{k}^{2}.\label{20.4}
\end{equation}
From Eqs. (\ref{consis}) and (\ref{10}) one obtains
\begin{equation}
\lambda\geq\mu_{k}^{2}+\left|\mu_{k}\right|\label{minmax}
\end{equation}
and
\[
\left|\mu_{k}+\frac{1}{2}\right|=\left|\mu_{k+1}-\frac{1}{2}\right|,
\]
whence
\begin{equation}
\mu_{k+1}=\mu_{k}+1,\label{20.5}
\end{equation}
which gives after iteration 
\begin{equation}
\mu_{k}=\mu_{0}+k,\quad k=0,1,2,\dots\label{20.6}
\end{equation}
where $k$ is the number of unit steps required to go from $\mu_{0}$
to $\mu_{_{k}}$, and $\mu_{0}$ denotes the minimum eigenvalue in
$\{\mu_{k}\}$. Introducing $\lambda=j(j+1),$ with $j>0$, Eq. \eqref{minmax}
rewrites as 
\begin{equation}
j+\frac{1}{2}\geq\begin{cases}
\mu_{k}+\frac{1}{2}, & \mu_{k}\geq0\\
\left|\mu_{k}-\frac{1}{2}\right|, & \mu_{k}\leq0
\end{cases}\label{20.7}
\end{equation}
which implies 
\[
-j\leq\mu_{k}\leq j.
\]
Since the number of steps required to go from $\mu_{0}=-j$ to $\mu_{M}=j$
must be an integer, the value of $j$ must be an integer or half integer,
and from \eqref{20.6} one gets
\begin{equation}
\mu_{k}=-j+k,\quad k=0,1,\dots2j.\label{20.10}
\end{equation}

The values obtained from (\ref{8}) and \eqref{20.4} for the coefficients,
namely
\[
|C_{k}|^{2}=\tfrac{1}{2}[j(j+1)-\mu_{k}(\mu_{k}+1)]=\tfrac{1}{2}[(j+\mu_{k}+1)(j-\mu_{k})],
\]
reproduce well-known results for the (squared modulus of the) matrix
elements of the ladder operators in the basis $\left\{ |k\rangle\right\} $.
Note that although the coefficients depend on both indices $j$ and
$k$, the dependence on the former is passive, so to say, meaning
that $j$ appears as a fixed parameter whilst $k$ refers to the state
on which the ladder operators act.
\selectlanguage{english}%

\subsection{The $N$-dimensional radial problem}

There exists a family of problems for which a factorization method
allows to construct appropriate state-independent operators $A$ and
$S$ that can be used to solve the eigenvalue problem with the help
of the consistency relations. These are $N$-dimensional radial problems,
i. e., central-force problems in $N$ dimensions ($N=1,2,3$) that
have been reduced to the radial variable only. Although the present
procedure has been applied to these problems in Ref. \cite{FeCa84},
we shall briefly present it here, along with some examples by way
of illustration.

To investigate the spectrum associated with the radial factor of such
problems, we introduce a couple of operators \cite{CiPal77,FeCa84}
that depend on a continuous, dimensionless variable $q$ as follows
(here $\varLambda$ is a real parameter)
\begin{equation}
B_{n\pm}=q^{n}\pm\left(\frac{-\Lambda}{q^{n}}+\frac{1}{q^{n-2}}\frac{d^{2}}{dq^{2}}\right),\label{138}
\end{equation}
with the properties \BA\label{Q+-} 
\begin{equation}
\left[q\frac{d}{dq},B_{n\pm}\right]=nB_{n\mp},\label{138aa}
\end{equation}

\begin{equation}
B_{n+}=2q^{n}-B_{n-}.\label{138ab}
\end{equation}
\CA This procedure is justified by the fact that for certain values
of the exponent $n$, $B_{n-}$ happens to have the form of an important
quantum-mechanical radial Hamiltonian, as will be seen below.

Careful inspection of Eqs. \eqref{Q+-} suggests to select \BA\label{Q}
\begin{equation}
\alpha=B_{n+},\label{13.139a}
\end{equation}
\begin{equation}
i\beta=q\frac{d}{dq}-\left(q\frac{d}{dq}\right)^{\dagger}=2q\frac{d}{dq}+(n-1)\mathbb{I},\label{13.139b}
\end{equation}
\CA so that after a few simple transformations one obtains \BA\label{SAQ}
\begin{equation}
S=\alpha^{2}+\beta^{2}=B_{n-}^{2}+(n^{2}-1-4\Lambda)\mathbb{I},\label{13.140a}
\end{equation}
\begin{equation}
A=i[\beta,\alpha]=2nB_{n-}.\label{13.140b}
\end{equation}
\CA This couple of equations indicates that the eigenvalues of the
operators $A$ and $S$ are related through the eigenvalues of the
operator $B_{n-}$, which are the ones of interest (in this case the
operator $P$ of Eq. \eqref{p} is just proportional to $A$). Denoting
these eigenvalues with $b_{k}$, we obtain from the consistency condition
\eqref{consis} 
\begin{equation}
b_{k+1}^{2}-b_{k}^{2}=2n(b_{k+1}+b_{k}).\label{13.141}
\end{equation}
Upon elimination of the common factor $b_{k+1}+b_{k}$ (assuming it
is different from zero, which is true in the cases of interest), this
equation simplifies into 
\[
b_{k+1}-b_{k}=2n,
\]
with solution 
\begin{equation}
b_{k}=b_{0}+2nk.\label{13.142}
\end{equation}
For $n>0$, which is the only case we shall study here, there is no
upper bound. Equation \eqref{rec} gives $2nb_{0}=b_{0}^{2}+n^{2}-1-4\Lambda$,
whence 
\begin{equation}
b_{0}=n\pm\sqrt{1+4\Lambda}\geq0.\label{13.143}
\end{equation}
We see that the proposed selection of the operators $\alpha,\beta$
has led to the general solution expressed in Eq. \eqref{13.142}.
The gist of the method was to obtain a nonlinear relation between
$s_{k}$ and $a_{k}$, with $a_{k}\propto b_{k}$, and use the consistency
condition to determine $b_{k}$.

We shall now apply this procedure to a couple of examples. 

\subsubsection{\textit{\emph{The isotropic $N$-dimensional harmonic oscillator}}}

The radial equation for the $N\text{-dimensional}$ isotropic harmonic
oscillator is 
\[
\left(-\frac{\hbar^{2}}{2m}\frac{d^{2}}{dr^{2}}+\frac{g}{2r^{2}}+\frac{1}{2}m\omega^{2}r^{2}\right)u=Eu,
\]
with $u=rR$, $R(r)$ being the radial part of the wave function,
and
\begin{equation}
g=\left[\tfrac{1}{4}(N-1)(N-3)+l(l+N-2)\right]\frac{\hbar^{2}}{m}.\label{13.143b}
\end{equation}
Notice that the angular momentum $l$ enters here as a fixed parameter.
The case $N=1$ has already been dealt with in Section 3.1.1, so we
shall consider here $N=2,3$. In terms of the dimensionless variable
$q=r/\alpha_{0}$ with $\alpha_{0}=\sqrt{\hbar/m\omega}$, the energy
becomes expressed in units of $\frac{1}{2}\hbar\omega$ and the Hamiltonian
coincides with $B_{n-}$ for $n=2$ \cite{FeCa84}, 
\begin{equation}
H=-\frac{d^{2}}{dq^{2}}+\frac{\Lambda}{q^{2}}+q^{2}=B_{2-}\label{13.144}
\end{equation}
where 
\begin{equation}
\Lambda=mg/\hbar^{2}=\tfrac{1}{4}(N-1)(N-3)+l(l+N-2).\label{13.145}
\end{equation}
Equation \eqref{13.143} with $n=2$ gives for $E_{0}$ 
\begin{equation}
E_{0}=\tfrac{1}{2}\hbar\omega b_{0}=\tfrac{1}{2}\hbar\omega\left(2\pm\sqrt{(N+2l-2)^{2}}\right).\label{13.145a}
\end{equation}
For $N=2,3$ and arbitrary $l$ the condition $E_{0}\geq0$ precludes
the negative sign, so that 
\[
E_{0}=\tfrac{1}{2}\hbar\omega(N+2l);\;N=2,3.
\]
 Combining this result with Eq. \eqref{13.142} for $n=2$, we obtain
for the energy corresponding to the radial state $k$ and a given
angular momentum $l$, 
\begin{equation}
E_{kl}=\hbar\omega(2k+l+N/2).\label{13.146}
\end{equation}
The subindex $l$ has been added to the energy eigenvalues to remind
us that the angular momentum eigenvalue enters as a parameter. 

\subsubsection{The hydrogen-like atom}

The radial Schrödinger equation for the hydrogen-like atom in $N$
dimensions is 
\begin{equation}
\left[-\frac{\hbar^{2}}{2m}\frac{d^{2}}{dr^{2}}+\frac{g}{2r^{2}}-\frac{Ze^{2}}{r}\right]u=Eu.\label{SHA10}
\end{equation}
Multiplying by $r$, in terms of the variable $q=\beta_{E}r$ with
$\beta_{E}=\sqrt{2m\left|E\right|}/\hbar$, the radial Hamiltonian
reduces after some simple rearrangements to the operator $B_{n-}$
for $n=1$ \cite{FeCa84},
\[
\left(-q\frac{d^{2}}{dq^{2}}+\frac{\Lambda}{q}+q\right)u=B_{1-}u=\frac{Ze^{2}}{\hbar}\sqrt{\frac{2m}{|E|}}u.
\]
with $\Lambda$ given by \eqref{13.145}. To determine $b_{0}=(Ze^{2}/\hbar)\sqrt{2m/|E_{0}|}$
we proceed as above, using \eqref{13.143} now with $n=1$, thus obtaining
$b_{0}=N-1+2l$ (the minus sign in Eq. \eqref{13.143} is ruled out
since $b_{0}\geq0$ for any $N$ and $l$). Using Eq. \eqref{13.142}
we get $b_{k}=N-1+2(l+k$), which combined with $b_{k}=(Ze^{2}/{\hbar)}\sqrt{2m/|E_{k}|}$
gives for the energy eigenvalues 
\begin{equation}
E_{kl}=-\frac{Z^{2}e^{4}m}{2\hbar^{2}(k+l+\frac{N-1}{2})^{2}}=-\frac{Z^{2}e^{4}m}{2\hbar^{2}\mathrm{n}^{2}}.\label{13.147}
\end{equation}
Note that the principal quantum number, denoted by $\mathrm{n}$ (roman),
$\text{n}=k+l+\frac{N-1}{2}$, is an integer in the 1D and 3D cases,
and a half-integer in the 2D case.

In Ref. \cite{Bo88}, the H atom problem is solved by means of a similar
procedure involving factorization of the Hamiltonian. Additional examples
of the central problem in several dimensions can be seen in Refs.
\cite{CaPl09,deLange92,PeMon80,FeCa84}. All these examples show that
even when central-force problems can be reduced for their solution
to one-dimensional (radial) problems regardless of their physical
dimension $N$, the energy spectrum does depend explicitly on the
number of dimensions.

\subsubsection{The Dirac H-atom}

The consistency conditions apply whenever the operators $S$ and $A$
can be written in the form of Eqs. \eqref{SAQ}, where $B_{n-}$ represents
the Hamiltonian operator expressed in the appropriate dynamical variable.
It is therefore possible to apply them also to a relativistic problem.
Let us select for this purpose the radial component of the hydrogen
atom as described by the Dirac theory. 

For a central problem, a standard---though elaborate---procedure
(see, e. g., Refs. \cite{Davy,La89}) allows to transform the Dirac
equation of first order for a four-component bi-spinor into a couple
of first-order equations, each for a two-component spinor and the
corresponding amplitudes. A further separation of the angular and
radial variables leads to a second-order equation corresponding to
the Klein-Gordon equation plus additional terms that represent corrections
generated by the coupling of the spin of the particle to the external
field. Finally, the angular and radial functions can be factored out.

For the present purpose let us consider the stationary case. The transformation
just described leads to the stationary radial equation for the hydrogen-like
atom,

\begin{equation}
\left(\nabla_{r}^{2}+\frac{\Lambda_{D}}{r^{2}}+\frac{2c_{E}}{r}-d_{E}\right)R=0,\label{DHA10}
\end{equation}
with \BA\label{DHA}

\begin{equation}
\nabla_{r}^{2}=\frac{1}{r}\frac{\partial^{2}}{\partial r^{2}}r,\;\Lambda_{D}=l(l+1)-\alpha^{2}Z^{2},\label{DHA12}
\end{equation}
\begin{equation}
d_{E}=\frac{m^{2}c^{2}}{\hbar^{2}}\left(1-\mathcal{E}^{2}\right),\;c_{E}=\frac{Zme^{2}}{\hbar^{2}}\mathcal{E},\label{DHA13}
\end{equation}
\CA where 
\begin{equation}
\mathcal{E}=\frac{mc^{2}+E}{mc^{2}}\label{DHA14}
\end{equation}
stands for the energy of the atom, including the rest energy, expressed
in units of the rest energy, and $\alpha=e^{2}/\left(\hbar c\right)$
is the fine-structure constant, $\alpha\simeq1/137$. In what follows
we shall limit the calculation to the hydrogen atom by making the
atomic number $Z=1$.

In terms of the function $u=rR$, Eq. \eqref{DHA10} acquires the
same structure as \eqref{SHA10}. They both correspond to the operator
$B_{1-}$, of course with different sets of parameters, and \eqref{DHA10}
is therefore equivalent to
\begin{equation}
\left(-q\frac{d^{2}}{dq^{2}}+\frac{\Lambda_{D}}{q}+q\right)u=B_{1-}u=\frac{2c_{E}}{d_{E}^{1/2}}u,\label{DHA16}
\end{equation}

\begin{equation}
\Lambda_{D}=l\left(l+1\right)-\alpha^{2},\qquad q=d_{E}^{1/2}r.\label{DHA17-1}
\end{equation}
Hence the solution for the Schrödinger atom can be translated to the
Dirac atom,
\begin{equation}
b_{k}=b_{0}+2k,\label{DHA20}
\end{equation}
with $b_{0}$ given according to Eq. \eqref{13.143} by $b_{0}=1+\sqrt{1+4\Lambda_{D}}.$
Here again the minus sign is ruled out since $b_{0}\geq0$ for any
$l$. From Eqs. \eqref{DHA} and \eqref{DHA20} we obtain after some
algebra

\begin{equation}
\mathcal{E}_{k}=\left[1+\frac{4\alpha^{2}}{\left(b_{0}+2k\right)^{2}}\right]^{-1/2},\label{eq:DHA22}
\end{equation}
which is the exact result for the energy of the Dirac H-atom in units
of $mc^{2}$ \cite{Davy}. 

\subsection{State-dependent individual operators; a simple example}

The harmonic oscillator, the angular momentum and the radial problems
just presented have an important feature in common: in all these cases,
the succesive eigenvalues of the relevant operator $P$ (with $P=H,J_{3},B_{n-}$,
respectively) are equally spaced, 
\begin{equation}
p_{k}=p_{0}+\gamma k,\label{P2}
\end{equation}
whence $p_{k+1}-p_{k}=\gamma$, where the value of the parameter $\gamma$
depends on the specific problem ($\gamma=\hbar\omega,1,2n$, respectively). 

For the harmonic oscillator and the angular momentum, the individual
ladder operators $\eta_{k}^{\dagger},\eta_{k}$ do not depend explicitly
on the state $k$. For state-dependent operators $\eta_{k}^{\dagger},\eta_{k}$
by contrast, using the algebraic method may not be the most convenient
procedure. This is clearly seen in Ref. \cite{YiBi87}, where the
eigenvalue problem of the H atom is solved by constructing a pair
of radial ladder operators, which turn out to contain the angular
momentum \emph{l} and thus give rise to \emph{l}-dependent commutators,
such as $\left[\eta_{l},\eta_{l\text{'}}^{\dagger}\right]=\left(l+l'+2\right)/r^{2}$.
Here we illustrate the point by resorting to another well-known elementary
textbook problem, namely the 1D infinite square well, with Hamiltonian
given by
\begin{equation}
H=\frac{p^{2}}{2m}=-\frac{\hbar^{2}}{2m}\frac{d^{2}}{dx^{2}};\;0\leq x\leq L.\label{P4}
\end{equation}
This problem has received considerable attention along the years,
and ladder operators have been obtained with the help of the \noun{susy}
technique \cite{CooKha95} and by means of alternative algebraic methods
(see \cite{JaTa13} and references therein). In this case the individual
operators $\eta_{k}^{\dagger},\eta_{k}$ turn out to be explicitly
state-dependent. In terms of the dimensionless variable $y=\left(\nicefrac{\pi}{L}\right)x$,
their expression \cite{DongMa02}, when acting on $\varphi_{k}(y)=\left\langle y\right.|k\rangle$,
is
\[
\eta_{i}^{\dagger}\varphi_{k}=\left(\cos y+\frac{1}{k}\sin y\frac{d}{dy}\right)\varphi_{k}\delta_{ik},
\]
\begin{equation}
\eta_{i}\varphi_{k}=\left(\cos y-\frac{1}{k}\sin y\frac{d}{dy}\right)\varphi_{k}\delta_{ik-1}.\label{P6}
\end{equation}
It can be readily confirmed that these operators satisfy Eqs. \eqref{13.122d}
by applying them to the (well-known) solutions for the problem, $\varphi_{k}(y)=\sqrt{2/L}\sin ky$.
In fact, given the trigonometric properties of the sine function it
is an easy matter to infer the structure of $\eta_{k}^{\dagger},\eta_{k}$.
With the selection of the coefficients appearing in Eqs. \eqref{eta}
as
\[
C_{k}=\sqrt{k(k+1)/2}\,,
\]
Eqs. \eqref{etaket} become 
\[
\eta^{\dagger}\varphi_{k}=\sqrt{k(k+1)/2}\,\varphi_{k+1},\;\eta\varphi_{k}=\sqrt{k(k-1)/2}\,\varphi_{k-1},
\]
and Eqs. \eqref{skak} give
\begin{equation}
s_{k}=k^{2},\;a_{k}=k.\label{P8}
\end{equation}
Thus the effect of operator $S$ is proportional to the effect of
the Hamiltonian, $H\varphi_{k}=E_{1}S\varphi_{k}$, with $E_{1}=\hbar^{2}\pi^{2}/2mL^{2}$,
as follows from Eq. \eqref{P4} applied to $\varphi_{k}(x)$. The
operator $A$, in turn, turns out to be isospectral to the harmonic
oscillator Hamiltonian, Eq. \eqref{eq:60.3}.

Clearly, when the eigenfunctions for a specific problem are known
in advance, as in the example just shown, the algebraic procedure
becomes unnecessary. However, in cases where the eigenfunctions are
not known or the Schrödinger equation is not so easily solved, construction
of the ladder operators, or else direct construction of a relevant
operator, may be the pathway to solve the eigenvalue problem with
the help of the algebraic consistency conditions, as in the examples
discussed above.

\subsection{An example from perturbation theory$ $}

As a demonstration of extra possibilities of the algebraic method
presented here, let us apply it to a traditional and simple exercise
of perturbation theory. We consider the 1D harmonic oscillator perturbed
by an interaction Hamiltonian of the form $\delta H=\epsilon x^{4},$
and calculate the energy spectrum to first order in $\epsilon.$ For
this purpose it is convenient to first establish some fundamental
relations.

We write the perturbed Hamiltonian as $H=H$$^{0}+\delta H,$ and
similarly $\alpha=$$\alpha^{0}+\delta\alpha,$ $\beta=\beta^{0}+\delta\beta,$
$S=S^{0}+\delta S,$ $A=A^{0}+\delta A$. Then to first order in $\epsilon$,
\BA\label{deltaSA}

\begin{equation}
\delta S=\alpha^{2}+\beta^{2}-S^{0}=\alpha^{0}\delta\alpha+\delta\alpha\alpha^{0}+\beta^{0}\delta\beta+\delta\beta\beta^{0},\label{13.174a}
\end{equation}
\begin{equation}
\delta A=-i\left[\alpha^{0},\delta\beta\right]-i\left[\delta\alpha,\beta^{0}\right].\label{13.174b}
\end{equation}
\CA We take $S=H$; since $S^{0}=H^{0},$ the eigenvalues of $S$
are, to first order, $s_{k}=E_{k}=E_{k}^{0}+\delta E_{k}$, and from
the consistency condition \eqref{consis} it follows, after cancelling
out the zero-order terms, that

\begin{equation}
\delta E_{k+1}-\delta E_{k}=\delta a_{k+1}+\delta a_{k},\label{13.175}
\end{equation}
with $\{\delta a_{k}\}$ the eigenvalues of $\delta A$. This is sufficient
for our present needs.

From the selection (see Eq. \eqref{134a})

\[
\alpha^{0}=\sqrt{\frac{m}{2}}\omega x,\quad\beta^{0}=\frac{1}{\sqrt{2m}}p
\]
it follows that $\delta\beta=0$ and $\delta\alpha$ becomes a function
of $\delta H,$ which being a function of $x$ commutes with $\alpha^{0}$.
Therefore,

\[
\delta S=\delta H=2\alpha^{0}\delta\alpha=\sqrt{2m}\omega x\delta\alpha,
\]
whence

\[
\delta\alpha=\left(\epsilon/\sqrt{2m}\omega\right)x^{3}.
\]
With these values, Eq. (\ref{13.174b}) becomes

\[
\delta A=-\frac{i\epsilon}{2m\omega}\left[x^{3},p\right]=\frac{3\hbar\epsilon}{2m\omega}x^{2},
\]
which introduced in Eq. (\ref{13.175}) gives 

\[
\delta E_{k+1}-\delta E_{k}=\frac{3\hbar\epsilon}{2m\omega}\left[\left\langle k+1\left|x^{2}\right|k+1\right\rangle +\left\langle k\left|x^{2}\right|k\right\rangle \right]=
\]
\begin{equation}
=\frac{3\hbar\epsilon}{2m\omega}\left[\frac{E_{k+1}^{0}}{m\omega^{2}}+\frac{E_{k}^{0}}{m\omega^{2}}\right]=\frac{3\hbar^{2}\epsilon}{m^{2}\omega^{2}}\left(k+1\right).\label{deltaak}
\end{equation}
An iteration gives

\[
\delta E_{k}=\delta E_{0}+\frac{3\hbar^{2}\epsilon}{m^{2}\omega^{2}}\sum_{i=1}^{k}i=\delta E_{0}+\frac{3\hbar^{2}\epsilon}{2m^{2}\omega^{2}}k\left(k+1\right).
\]
To fix the value of $\delta E_{0}$ we recall that $s_{0}=a_{0}$,
which gives

\[
s_{0}=\delta E_{0}=\delta a_{0}=\frac{3\hbar\epsilon}{2m\omega}\left\langle 0\left|x^{2}\right|0\right\rangle =\frac{3\hbar^{2}\epsilon}{4m^{2}\omega^{2}}.
\]
The first-order correction to the energy eigenvalue is therefore

\begin{equation}
\delta E_{k}=\frac{3\hbar^{2}\epsilon}{2m^{2}\omega^{2}}\left(k^{2}+k+\frac{1}{2}\right).\label{eq:13.178}
\end{equation}
For a more general study of the oscillator subject to a polynomial
perturbation in $x$ and $p$, see Ref. \cite{BoDa18}.

\subsubsection*{Acknowledgments}

The authors are grateful to the referees for their valuable comments.
Partial support from DGAPA-UNAM through project PAPIIT IN113720 is
acknowledged.

\end{document}